\documentclass[12pt,nohyper,letterpaper]{JHEP3}         
\usepackage{amssymb,amsfonts}

\usepackage{epsfig}

\def\be{\begin{eqnarray}}
\def\ee{\end{eqnarray}}
\newcommand{\nn}{\nonumber}
\newcommand\para{\paragraph{}}
\newcommand{\ft}[2]{{\textstyle\frac{#1}{#2}}}
\newcommand{\eqn}[1]{(\ref{#1})}

\def\Dslash{\,\,{\raise.15ex\hbox{/}\mkern-12mu D}}
\def\Dbarslash{\,\,{\raise.15ex\hbox{/}\mkern-12mu {\bar D}}}
\def\delslash{\,\,{\raise.15ex\hbox{/}\mkern-9mu \partial}}
\def\delbarslash{\,\,{\raise.15ex\hbox{/}\mkern-9mu {\bar\partial}}}
\def\pslash{\,\,{\raise.15ex\hbox{/}\mkern-9mu p}}
\def\calDslash{\,\,{\raise.15ex\hbox{/}\mkern-12mu {\cal D}}}

\newcommand{\sign}{{\rm sign}}

\newcommand{\Tr}{{\rm Tr}}

\def\v{\varphi}
\def\tv{\tilde{\varphi}}

\def\lae{\mathrel{\mathop{\smash{\lower .5 ex \hbox{$\stackrel<\sim$}}}}}
\def\lae{\mathrel{\mathop{\smash{\lower .5 ex \hbox{$\stackrel>\sim$}}}}}

\title{Instantons, Fermions and Chern-Simons Terms}
\author{Benjamin Collie and David Tong \\
Department of Applied Mathematics and Theoretical Physics, \\
University of Cambridge, UK\\
{\tt b.p.collie, d.tong@damtp.cam.ac.uk}}

\abstract{In five spacetime dimensions, instantons are finite
energy, solitonic particles. We describe the dynamics of these
objects in the presence of a Chern-Simons interaction. For $U(N)$
instantons, we show that the 5d Chern-Simons term induces a corresponding
Chern-Simons
term in the ADHM quantum mechanics. For $SU(N)$ instantons, we
provide a description in terms of geodesic motion on the instanton
moduli space, modified by the presence of a magnetic field. We
show that this magnetic field is equal to the first Chern
character of an index bundle. All of these results are derived by
a simple method which follows the fate of zero modes as fermions
are introduced, made heavy, and subsequently integrated out.}

\begin{document}
\pagestyle{plain} \setcounter{page}{1}
\newcounter{bean}
\baselineskip16pt

\setcounter{section}{1} \setcounter{equation}{0}

\section*{1. Introduction}

In five spacetime dimensions, the self-dual instanton solutions of
Yang-Mills theory are finite energy, solitonic particles. Their
dynamics is described by quantum mechanics on the instanton moduli
space ${\cal M}$. The purpose of this paper is to explain how this
dynamics is altered by the presence of a 5d Chern-Simons (CS)
term.

\para
This topic has received some interest of late, motivated in part
by an old observation of Atiyah and Manton which relates
instantons to Skyrmions \cite{am}. This idea has recently found a
dynamical realization within the context of holographic QCD
\cite{sugimoto,sugi2}\footnote{There are also a number of further
proposals which provide different dynamical realizations of the
relationship between 5d instantons and Skyrmions
\cite{hill,son,me}.}. In this framework, a $U(N_f)$ Yang-Mills
theory in a five-dimensional warped background reduces at
low-energies to the Skyrme model with $N_f$ flavours. The 5d
theory includes CS interactions which play a key role in the story
and, for $N_f \geq 3$, give rise to the WZW term of the Skyrme
model. Baryons in this model arise from instantons in 5d and have
been studied in a number of papers [7-11].

\para
Here we study the dynamics of instantons  in flat five-dimensional
Minkowski space. To derive the effects of the CS interactions on
the instanton dynamics, we use a simple method which has its roots
in the earlier, classic work of \cite{gw,cw}. We examine the
effects of massive fermions which, once integrated out, generate
the CS terms of interest. By following the fate of the fermi zero
modes, one arrives at a simple description of the instanton
dynamics in terms of geodesic motion, modified by the presence of
a background magnetic field on ${\cal M}$. In a companion paper
\cite{3dcs}, we use this technique to describe the dynamics
of vortices in three-dimensional CS theories.

\para
We primarily focus on the dynamics of BPS instantons in $d=4+1$
$U(N)$ theories with ${\cal N}=1$ supersymmetry. In the absence of
CS terms, a simple and convenient description of the instanton
dynamics is provided by the ADHM quantum mechanics
\cite{adhm,doug}. Our main result is that the 5d CS term has a
very simple effect in the ADHM language: it induces a
corresponding CS term in the quantum mechanics of the type
discussed previously in \cite{eva}.

\para
We start in Section 2 by reviewing some aspects of the 5d theory,
including the instanton solutions and manner in which fermions
induce CS interactions. In Section 3, we turn to the effective
theory of the instanton dynamics, working firstly in the ADHM
formalism. We show how fermions induce CS terms in the quantum
mechanics. We end with a discussion of $SU(N)$ instantons (as
opposed to $U(N)$ instantons), where we provide a more geometric
description of the instanton dynamics. We show that the magnetic
field over ${\cal M}$ is given by the first Chern character of the
bundle of fermi zero modes. Related results were obtained previously
in the context of counting instantons in 5d Super Yang-Mills with CS terms
\cite{tachikawa}.

\para
{\bf Note added:} In the final stages of preparation of this paper,
a preprint appeared on the arXiv which contains substantial overlap
with our work \cite{damn}.

\setcounter{section}{2}
\section*{2.  Instantons in 5d Chern-Simons Theories}

In this section we consider the five-dimensional theory with
${\cal N}=1$ supersymmetry (i.e. 8 supercharges) and gauge group
$G=U(N)$. The vector multiplet contains a gauge field $A_\mu$, a
real adjoint scalar $\phi$ and an adjoint Dirac fermion $\lambda$.
In the absence of a Chern-Simons interaction, the bosonic part of
the Lagrangian is given by the familiar expression\footnote{We choose
anti-hermitian generators of the Lie algebra satisfying
$[T_a,T_b]=f_{ab}^cT_c$, normalized such that
$\Tr(T_aT_b)=-\ft12\delta_{ab}$. The field strength is
$F_{\mu\nu}^a=\partial_\mu A_\nu^a-\partial_\nu A_\mu^a +
f_{bc}^aA_\mu^bA_\nu^c$ and the covariant derivative is ${\cal
D}_\mu\phi^a=\partial_\mu\phi^a+f_{bc}^aA_\mu^b\phi_c$. We define
$F_{\mu\nu}=F_{\mu\nu}^aT_a$, etc.}
\be {\cal L}_{YM} =
\frac{1}{e^2}\,\delta_{ab}\left(-\frac{1}{4}F^a_{\mu\nu}F^{b\mu\nu}
-  \frac{1}{2}{\cal D}_\mu\phi^a{\cal
D}^{\mu}\phi^b\right)\label{ymlag}\ee
When $\langle \phi\rangle = 0$, the gauge symmetry is unbroken and
this theory admits the familiar self-dual instanton solutions,
\be F_{ij} = {}^\star F_{ij}\ \ \ \ \ \ \ \ \ i,j=1,\ldots,4
\label{sd}\ee
In five spacetime dimensions, these are localized solitonic
particles. They carry charge $k\in {\bf Z}$ under the global
$U(1)_I$ symmetry with conserved current
$J\sim{}^\star\,\Tr\,F\wedge F$ \cite{seiberg}, and have mass
\be M_{\rm inst} = \frac{8\pi^2 k}{e^2}\label{mass}\ee
The self-duality equations \eqn{sd} have a moduli space ${\cal M}$
of solutions, of real dimension ${\rm dim}({\cal M})=4kN$. Away
from singularities, the low-energy dynamics is well described by
geodesic motion on ${\cal M}$ and this provides a good starting
point for understanding the quantization of these objects. There
is an elegant and convenient way of packaging this dynamics in
terms of the ADHM matrix model which we shall review in Section 3.

\para
Let us now add to \eqn{ymlag} the supersymmetric Chern-Simons
Lagrangian \cite{seiberg,ims}
\be {\cal L}_{CS} = \frac{c}{4\pi^2}\,d_{abc}\phi^a\left( -
\frac{1}{4}F^b_{\mu\nu}F^{c\mu\nu} - \frac{1}{2}{\cal
D}_\mu\phi^b{\cal D}^\mu\phi^c  \right) +
\frac{c}{24\pi^2}\,\omega(A) \label{cslag}\ee
Here $d_{abc}$ is the totally symmetric tensor
\be d_{abc} = \ft12 \Tr\,( T^a\{T^b,T^c\})\ .\ee
The last term $\omega(A)$ is the Chern-Simons five-form
\be \omega(A) =
\epsilon^{\mu\nu\rho\sigma\lambda}\Tr\left(A_\mu\partial_\nu
A_\rho\partial_\sigma A_\lambda -\frac{3i}{2} A_\mu A_\nu A_\rho
\partial_\sigma A_\lambda -\frac{3}{5} A_\mu A_\nu A_\rho A_\sigma
A_\lambda\right)\label{cs}\ \ee
The first and second terms in \eqn{cslag} are dictated by
supersymmetry \cite{seiberg,ims}. In the presence of the CS term,
invariance of the partition function under large gauge
transformations imposes the quantization requirement $c \in {\bf
Z}$.

\para
The CS term couples the gauge field to the instanton current $J$,
telling us that instantons now carry non-Abelian gauge
charge\footnote{When $\langle\phi\rangle\neq0$, instantons can
also carry Abelian gauge charge, even in the absence of a
Chern-Simons term \cite{dyonic}.} \cite{seiberg}. This ensures
that the self-dual configurations \eqn{sd} now source $A_0$ and
$\phi$. We work in static gauge $\partial_0=0$, and set
$A_0=\phi$. It can be shown that the equations of motion are
satisfied if the self-dual equations are supplemented by Gauss'
law, which can be written as \cite{seok}:
\be{\cal D}^2 \left(\frac{\phi^a}{e^2} + \frac{c}{8\pi^2}
d_{abc}\phi^b\phi^c\right) =
\frac{c}{16\pi^2}d_{abc}F_{ij}^bF^c_{ij}\ee
Solutions to this equation were studied using the ADHM technique in
\cite{seok}. There exists a unique solution for each self-dual
configuration. The mass \eqn{mass} of the
instanton remains unchanged in the presence of a CS interaction. We stress
that the existence of the
$\phi$ field, with interactions \eqn{cslag}, plays a crucial role
in keeping the mass of the instanton unchanged.
Indeed, if one
considers the $U(N)$ CS theory without the $\phi$ field then no
solution exists: the CS term induces an electric charge, forcing
the instanton to large size \cite{ssbaryon}. We will return to
this case at the end of Section 3.

\para
The question that we want to ask is: how do the instantons move in
the presence of a CS interaction? Our strategy for answering this
question is simple, but indirect. We will replace the Chern-Simons
interactions \eqn{cslag} with something that we understand well:
fermions.

\subsection*{Integrating Out Hypermultiplet Fermions}

Instead of working directly with the Chern-Simons Lagrangian
\eqn{cslag}, we set $c_{\rm classical}=0$ and introduce
$N_f$ hypermultiplets, each of mass $m$, in the fundamental
representation of $G$. Each of these hypermultiplets contains two
complex scalars and a single Dirac fermion $\Psi$, which satisfies
\be \calDslash\Psi-(i\phi+m)\Psi=0 \label{dirac}\ee
(The unfamiliar factor of $i$ in front of $\phi=\phi^aT^a$ arises
because we are working with anti-Hermitian generators $T^a$).
Integrating out these fermions induces the interaction
\eqn{cslag}, with coefficient \cite{niemi,5dwit}
\be c = -\frac{N_f}{2}\,\sign(m) \label{cnf}\ee
(Here we've assumed $\Tr\phi =0$ in vacuum; otherwise it is
$\sign(Nm+i\Tr\phi)$ which is the relevant quantity). As well as
the induced Chern-Simons term, there is also a finite
renormalization of the gauge coupling \cite{ims}
\be \frac{1}{e^2} \longrightarrow \frac{1}{e^2_{\rm ren}} =
\frac{1}{e^2} + \frac{c}{4\pi^2}\,m\label{e2}\ee
Notice that this shifts the instanton mass $M_{\rm inst}
\rightarrow M_{\rm inst} - N_f |m|$, an effect first pointed out
in \cite{brane}.

\para
All other effects of  integrating out the hypermultiplets are
suppressed by powers of $1/m$. By taking the limit $m\rightarrow
\infty$, keeping $e^2_{\rm ren}$ fixed, we ensure that the
resulting theory is the same as that described by the Lagrangians
\eqn{ymlag} and \eqn{cslag}. We will now study the effects of the
fermions on the instanton dynamics.

\setcounter{section}{3}
\section*{3. Instanton Dynamics}

The low-energy dynamics of instantons in a $U(N)$ gauge theory is
succinctly captured in the ADHM matrix model\footnote{The fact
that the ADHM matrix model describes $U(N)$ instantons, rather
than $SU(N)$ instantons, is clear from the brane picture
\cite{doug} and also from the generalization to non-commutative
backgrounds \cite{noncom}. While this distinction is irrelevant
for pure Yang-Mills theory, it becomes important in the presence
of fundamental matter or, as stressed in \cite{ssbaryon},
Chern-Simons interactions.} \cite{adhm,doug}. For a detailed
review, see \cite{dhvm}. In this section, we first review the
ADHM formalism for instanton dynamics in the Yang-Mills theory
described by the Lagrangian \eqn{ymlag}. We then derive the
modification of the dynamics once we add the CS interaction
\eqn{cslag}.

\para
The dynamics of $k$ instantons in $d=4+1$ dimensional $U(N)$
Yang-Mills theory can be formulated in terms of a $U(k)$ gauge
theory on the $d=0+1$ instanton worldline. When the 5d theory has
${\cal N}=1$ supersymmetry (i.e. 8 supercharges), the instanton
worldline theory has ${\cal N}=(0,4)$ supersymmetry\footnote{This
arises from the dimensional reduction of the chiral ${\cal
N}=(0,4)$ superalgebra in two dimensions. It is sometimes called
$N=4B$ supersymmetry to distinguish it from the ${\cal N}=(2,2)$
superalgebra.}. One introduces an auxiliary $U(k)$ vector
multiplet which contains a $U(k)$ Hermitian worldline gauge field
$\alpha$ together with a single real adjoint scalar $\sigma$. The
dynamical fields, which parameterize the instanton moduli space,
include two complex $k\times k$ matrices $W$ and $Z$, each of
which transforms in the adjoint of the $U(k)$ gauge group. There
are a further $N$ $k$-vectors $\varphi_i$, $i=1,\dots, N$,
transforming in the fundamental representation of $U(k)$, and  $N$
$k$-vectors $\tilde{\varphi}_i$ transforming in the
anti-fundamental representation.

\para
The low-energy dynamics of instantons is described by the Lagrangian                                        %
\be L = && \Tr\left( |{\cal D}_t Z|^2 + |{\cal D}_tW|^2
+ |{\cal D}_t\v_i|^2 + |{\cal D}_t\tv_i|^2 \right) \nn\\
&& \ \ \ - \Tr\left(|[\sigma,Z]|^2+|[\sigma,W]|^2 \right)-
\v^\dagger_i\sigma^2\v_i-\tv_i\sigma^2\tv_i^\dagger \label{mm}\ee
where ${\cal D}_tZ=\dot{Z}-i[\alpha,Z]$, and similar for $W$,
while ${\cal D}_t\varphi_i=\dot{\varphi}_i-i\alpha\varphi_i$ and
${\cal
D}_t\tilde{\varphi}_i=\dot{\tilde{\varphi}}_i+i\tilde{\varphi}_i\alpha$.
 The fields are  subject to constraints which arise from the
D and F-terms in the vector multiplet. They are given by
\be [Z,Z^\dagger] + [W,W^\dagger] + \v_i\v_i^\dagger -
\tv_i{}^\dagger\tv_i &=&0  \nn\\
{[Z,W]} + \v_i\tv_i &=& 0 \label{adhm}\ee
After quotienting by the gauge action $U\in U(k)$
\be Z\rightarrow UZU^\dagger\ \ \ ,\ \ \  W\rightarrow
UWU^\dagger\ \ \ ,\ \ \ \v_i\rightarrow U\v_i\ \ \ ,\ \ \
\tv_i\rightarrow \tv_iU^\dagger\ee
the scalars provide coordinates on the instanton moduli space
${\cal M}$. Given a suitable parametrization of the constraints
\eqn{adhm}, the first line in \eqn{mm} reduces to the
hyperK\"ahler metric on ${\cal M}$.

\para
The auxiliary field $\sigma$ that appears on the second line of
\eqn{mm}  plays no role in the purely bosonic theory (at least if
$\langle\phi\rangle=0$). However the ${\cal N}=(0,4)$ ADHM
Lagrangian also comes with a number of Grassmann valued fields.
These describe the zero modes of the adjoint Dirac fermion
$\lambda$ and live in the tangent bundle of ${\cal M}$. They won't
play an important role in our story, so we do not describe them
here. When they are included, integrating out $\sigma$ induces
four-fermi terms in the instanton action.

\subsection*{Integrating out Hypermultiplet Fermions}

Let us now examine the effect of the supplementary fermions $\Psi$
 that we introduced solely for the purpose of inducing
Chern-Simons interactions in 5d. We start by studying solutions to
the Dirac equation \eqn{dirac} in the background of a self-dual
instanton $F={}^\star F$ configuration. Of the full spectrum of
solutions, only the zero modes will prove important. We discuss
these first, and then explain why the non-zero modes are
unimportant. We choose the basis of $4\times 4$ gamma matrices
\be \gamma^0=\left(\begin{array}{cc} -i & 0 \\ 0 & +i
\end{array}\right)\ \ ,\ \ \gamma^i=\left(\begin{array}{cc} 0 &
\sigma^i \\ \sigma^i & 0 \end{array}\right) \ \ \ ,\ \ \
\gamma^4=\left(\begin{array}{cc} 0 & i
\\ -i & 0 \end{array}\right) \ee
with $\sigma^i$, $i=1,2,3$, the Pauli matrices. For each $\Psi$,
the ``zero mode" solutions arise from the ansatz $\Psi^T=
(e^{+imt}\psi_+,e^{-imt}\psi_-)$ where $\psi_\pm$ are two-component spinors that satisfy
\be \Dslash\psi_- \equiv \sigma_i {\cal D}_i\psi_- + i {\cal
D}_4\psi_- &=& 0 \nn\\ \bar{\Dslash}\psi_+ \equiv \sigma_i {\cal
D}_i\psi_+ - i {\cal D}_4\psi_+ &=& 0\ee
In the self-dual instanton background, standard index theorems
state that $\Dslash$ has $k$ complex zero modes, while
$\bar{\Dslash}$ has none. We introduce Grassmann collective
coordinates $\xi_-^l(t)$ for these modes, with $l=1,\ldots,k$, so $\xi_-$ transforms
in the ${\bf k}$ representation of $U(k)$. The worldline Lagrangian governing
the dynamics is simply
\be L_{\rm fermi} = \bar{\xi}_-(iD_t+\sigma+m)\xi_- \label{xi}\ee
There are no further constraints on $\xi_-$. (See, for example Section VI of
\cite{dhvm}, where $\xi_-$ is referred to as ${\cal K}$). There are further
four-fermi interactions, coupling $\xi_-$ with the vector multiplet fermi zero modes,
but they will not be important here\footnote{There is a subtlety with supersymmetry: the zero modes $\xi_-^l(t)$ live in an ${\cal N}=(0,4)$ fermi multiplet which naively includes twice as many zero modes. In 4d Euclidean space, these extra modes arise because one
relaxes the relationship between spinors and their conjugates. However, in 5d Minkowski space, this doubling does not occur. The resolution to this discrepancy seemingly follows from correctly
accounting for the constraints imposed on pseudo-Majorana spinors in $d=4+1$.}.

\para
We now come to quantize these Grassmann collective coordinates.
Firstly note that, although we have been referring to them as
``zero modes", this is a slight misnomer. As is clear from
\eqn{xi}, they are excited at a cost of energy equal to $|m|$.
This is unsurprising, since they arose from the 5d fermion $\Psi$
with mass $m$. Quantization gives rise to the usual ordering
ambiguity. Comparison with renormalization of $e^2$ suggests that
the correct quantization takes the ground state to have energy
$-|m|$ \cite{brane}.

\para
Our interest here lies in the effect of these zero modes on the
low-energy bosonic dynamics of the instantons. To see this effect, we
integrate out $\xi_-$ at one loop. A
standard computation of the determinants, given in  \cite{eva},
shows that each of these fermions contributes a term,
\be L_{\rm eff} = \log\left[\frac{\det (i\partial_t +
\alpha+\sigma+m)}{\det(i\partial_t+m)}\right] =\frac{1}{2}
\sign(m)\, \Tr(\alpha + \sigma)\label{yes}\ee
The $\Tr(\alpha)$ term is a quantum mechanical Chern-Simons
coupling.
The $\Tr(\sigma)$ term is a linear potential for the auxiliary
field; we will see the meaning of this shortly.  Integrating out
all $N_f$ hypermultiplet zero modes, and comparing to \eqn{cnf},
we arrive a description of the low-energy dynamics
of instantons in the theory with 5d CS interactions \eqn{cslag} in terms of an  ADHM CS matrix model,
\be L = && \Tr\left(|{\cal D}_t Z|^2 + |{\cal D}_tW|^2
+ |{\cal D}_t\v_i|^2 + |{\cal D}_t\tv_i|^2 -c (\alpha+\sigma) \right) \nn\\
&& \ \ \ -\Tr\left(|[\sigma,Z]|^2+|[\sigma,W]|^2\right)-
\v^\dagger_i\sigma^2\v_i-\tv_i\sigma^2\tv_i^\dagger
\label{csmm}\ee
This equation, which was also derived in \cite{damn}, is the main result of
this paper. Note that invariance under large gauge transformations requires
that the coefficient $c$ is integer valued, which coincides with the
requirement in the 5d theory.

\subsubsection*{Why Only Zero Modes Matter}

In deriving the Lagrangian \eqn{csmm}, we have integrated out only
the zero modes on the instanton worldline, while ignoring the
infinite tower of higher solutions to the Dirac equation. We now
show that this is consistent. The key point is that higher
excitations of fermions come in pairs, with energy $\pm E$:
\be \left(\begin{array}{cc} 0 & \Dslash  \\ \bar{\Dslash}
& 0 \end{array}\right)\left(\begin{array}{c} \psi_+ \\
\psi_- \end{array}\right) =  E
\left(\begin{array}{c} \psi_+ \\
\psi_- \end{array}\right) \ \ \Rightarrow\ \ \
\left(\begin{array}{cc} 0 & \Dslash  \\ \bar{\Dslash}
& 0 \end{array}\right)\left(\begin{array}{c} \psi_+ \\
-\psi_- \end{array}\right) =  -E
\left(\begin{array}{c} \psi_+ \\
-\psi_- \end{array}\right)\nn\ee
Contributions to the  Chern-Simons term on the instanton worldline
cancel between each pair. To see this, we write the general
eigenfunction as ${\Psi}^T = ({\psi}_-\zeta_-, {\psi}_+\zeta_+)$
and promote $\zeta_\pm$ to time-dependent Grassmann fields. The
action for these objects is schematically
\be L_{\rm non-zero-modes} = \bar{\zeta}_+(iD_t-m)\zeta_+ +
{\bar{\zeta}}_-(iD_t+m){\zeta}_- + E({\bar{\zeta}}_+{\zeta}_- +
\bar{\zeta}_-\zeta_+)\ee
which is schematic only in the sense that we have dropped overall
coefficients that arise from the overlap of the eigenfunctions.
Integrating out the non-zero modes now gives us a determinant of
the form
\be \det\left(\begin{array}{cc} iD_t -m & E \\ E &
iD_t+m\end{array}\right)= \det(iD_t +\sqrt{m^2+E^2})\det(iD_t -
\sqrt{m^2 + E^2})\ \ \ \ \ \ee
We see that the effective mass of these objects is
$\pm\sqrt{m^2+E^2}$, leading to a cancellation due to the presence
of the  $\sign(m)$ term in \eqn{yes}. In the limit $m\rightarrow
\infty$, these non-zero modes leave no trace of their existence on
the vortex dynamics.

\subsubsection*{An Example: A Single Instanton}

Let us study the dynamics \eqn{csmm} for a single $k=1$ instanton.
In this case $Z$ and $W$ describe the trivial centre of mass
motion and may be ignored. The remaining $4(N-1)$ collective
coordinates describe the scale size $\rho$ and the $U(N)$
orientation of the instanton. In the ADHM formalism, these are
captured by $2N$ complex scalars $\varphi_i$ and
$\tilde{\varphi}_i$, $i=1\ldots,N$, subject to the constraints
arising from \eqn{adhm}
\be \sum_i |\varphi_i|^2-|\tilde{\varphi}_i|^2 = \sum_i
\varphi_i\varphi_i = 0\ee
and modulo the $U(1)$ gauge action: $\varphi_i\rightarrow
e^{i\theta}\varphi_i$ and $\tilde{\varphi}_i\rightarrow
e^{-i\theta}\tilde{\varphi}_i$. These complex scalars define the moduli space
${\cal M}\cong T^\star{\bf CP}^{N-1}$, the cotangent bundle of
${\bf CP}^{N-1}$, with vanishing K\"ahler class. The singularity
at the origin of ${\cal M}$ corresponds to the singular, small
instanton. Away from the singularity, the scale size $\rho$ of the
instanton is identified as
\be \sum_i |\varphi_i|^2 + |\tilde{\varphi}_i|^2 =
\frac{16\pi^2}{e^2}\rho^2 \ee
To illustrate the various terms in the Lagrangian \eqn{csmm}, let
us first ignore the auxiliary field $\sigma$ and concentrate only
on the kinetic terms. They are given by,
\be L = |{\cal D}_t\varphi_i|^2 + |{\cal D}_t\tilde{\varphi}_i|^2									 %
- c \alpha = |\dot{\varphi}_i|^2 + |\dot{\tilde{\varphi}}_i|^2
-\alpha (j+c) + \frac{16\pi^2}{e^2}\alpha^2\rho^2 \ee
where
$j=i\sum_i[(\dot{\varphi}_i^\dagger\varphi_i-\varphi_i^\dagger\dot{\varphi}_i)
-
(\dot{\tilde{\varphi}}_i^\dagger\tilde{\varphi}_i-\tilde{\varphi}_i^\dagger
\dot{\tilde{\varphi}}_i)]$ is the $U(1)$ gauge current. To perform the
$U(1)$ quotient, we  simply integrate out the gauge field
$\alpha$. The resulting Lagrangian is given by,
\be L =  \left[|\dot{\varphi}_i|^2 + |\dot{\tilde{\varphi}}_i|^2 -									 %
\frac{e^2j^2}{64\pi^2\rho^2}\right]  - \frac{e^2}{32\pi^2}
\frac{cj}{\rho^2} - \frac{e^2c^2}{64\pi^2\rho^2}\label{last}\ee
The first three terms in square brackets are quadratic in
velocities: they implicitly define the metric on the instanton
moduli space. (An explicit parametrization of the constraints, and
hence the metric, was given in the $N=2$ case in \cite{mekim}).
The fourth term, linear in velocity, is the induced magnetic field
on the moduli space. Yet there is also a potential term, of order
$c^2$, which forces the instanton to large size $\rho\rightarrow
\infty$. This term is cancelled by the linear potential for
$\sigma$. Indeed, solving \eqn{csmm} for $\sigma$ gives,
\be \sigma = \frac{-c}{2\left(|\varphi_i|^2+|\tilde{\varphi}_i|^2\right)}\ee									 %
which, after substitution, cancels the last term in \eqn{last}. It
is simple to check that a similar phenomenon happens for the
general non-Abelian matrix model \eqn{csmm}, with the auxiliary
field $\sigma$ cancelling an effective potential on ${\cal M}$
induced by the CS term $c\alpha$.

\para
We note that the interplay between the potential generated by the
CS term and the linear $\sigma$ potential in the instanton theory
mimics a similar interplay in 5d. Without the scalar field $\phi$,
the $U(N)$ CS term generates a potential for the instanton,
pushing it to large size. This potential is cancelled by the
scalar field $\phi$ which adjusts so that the instanton once again
saturates the BPS bound \eqn{mass} for all scale sizes $\rho$.
Given this correspondence, one might wonder whether the $U(N)$
theory without $\phi$ is captured by the ADHM model without
$\sigma$. However, this does not appear to be the case. One can
compute the potential on the instanton moduli space generated by
the CS coupling (see, for example \cite{ssbaryon}).  In the $U(2)$
theory, the functional form of this potential agrees with that in \eqn{last},
but the numerical
coefficient differs.

%
%

\subsubsection*{A Variation on the Theme: $SU(N)$ Instantons}

The ADHM matrix model describes instantons in  $U(N)$ gauge theory
with the scalar field $\phi$. For instantons in pure $SU(N)$
Yang-Mills theory (without the scalar field $\phi$) we now derive
a more geometrical, albeit somewhat implicit, description of the
dynamics. Our strategy is as before: we introduce a fermion of
mass $m$ which, when integrated out, induces a 5d CS term. This
fermion also gives rise to a number of instanton zero modes. These
define a bundle over the instanton moduli space ${\cal M}$,
commonly referred to as the index bundle \cite{mansch}. As we
adiabatically change the instanton configuration, and hence move
in ${\cal M}$, the zero modes undergo a holonomy described by a
connection $\omega$ over ${\cal M}$. We denote the Grassmann
valued coordinates of the fibre as $\xi^m$. The appropriate
covariant derivative for these collective coordinates is given by
\be {\cal D}_t \xi^m = \partial_t\xi^m+(\omega_a)^m_{\
n}\dot{X}^a\xi^n\ee
where $X^a$ are coordinates on ${\cal M}$. Comparing with the ADHM
description \eqn{xi}, we see that the index bundle in that case
arose because the Grassmann collective coordinates were charged
under the auxiliary $U(k)$ gauge field. In the present case, the
description of the index bundle is more geometric. As shown in
\cite{3dcs}, integrating out the $\xi^m$ induces a sigma-model CS
term,
\be L_{CS} = \ft12\,\sign(m)\Tr(\omega_a)\dot{X}^a\ee
This is equivalent to a magnetic field $B=\Tr\,d\omega$ over the
instanton moduli space. This is simply the first Chern character
of the index bundle. We conclude that the low-energy dynamics of
instantons is described by geodesic motion on ${\cal M}$, modified
by the presence of this magnetic field.

\para
Let's look at some examples of this procedure. Consider firstly
the effect of adding a massive fermion $\lambda$ in the adjoint
representation of $SU(N)$. This does not contribute to a 5d CS term
because $d_{abc}=0$ for real representations. We can also see this
from the perspective of the instanton moduli space. For an adjoint
fermion, index theorems
tell us that there exist $2kN$ complex fermi zero modes. These can be
mapped to the bosonic zero modes of
the instanton, which ensures that they live in the tangent bundle
of ${\cal M}$. As explained in \cite{3dcs}, integrating out these
modes gives rise to a magnetic field on ${\cal M}$ which is equal
to the first Chern character. Yet the instanton moduli space is
hyperK\"ahler, and hence Calabi-Yau, and its first Chern character
vanishes. We thus see from both 5d and 1d points of view that
integrating out adjoint fermions does not affect the instanton
dynamics.

\para
Let us now consider a massive fermion $\Psi$ in the fundamental
representation of $SU(N)$. For $N\geq 3$, integrating out this
fermion in 5d induces a CS term with $c=-\ft12\sign(m)$. The
instanton picks up $k$ fermi zero modes which live in an index
bundle with a $U(k)$ connection. Integrating out these zero modes
in the worldline theory induces a magnetic field over the moduli space
${\cal M}$, equal to the first Chern character of the index
bundle. The low-energy dynamics of the instantons is described by
geodesic motion on ${\cal M}$ in the presence of this magnetic
field.

\para
For $SU(2)$, no CS term is generated in 5d because $d_{abc}=0$ for
all representations. Once again, we can see this from the
perspective of instanton zero modes. The index bundle for $SU(2)$
instantons has a $O(k)$ connection, rather than a $U(k)$
connection \cite{mansch}, ensuring that the first Chern class vanishes
and no magnetic field is generated on ${\cal M}$.

\section*{Acknowledgement}
We would like to thank Ben Allanach, Seok Kim, Ki-Myeong Lee,
Sungjay Lee  and Piljin Yi for useful
discussions.  BC is supported by an STFC studentship.  DT is supported by the Royal Society.

\end{document}